\documentstyle[manuscript,aps,prbbib,graphicx]{revtex}

\tighten
\newcommand{\be}{\begin{equation}}
\newcommand{\ee}{\end{equation}}
\newcommand{\bea}{\begin{eqnarray}}
\newcommand{\eea}{\end{eqnarray}}
\newcommand{\bd}{\begin{displaymath}}
\newcommand{\ed}{\end{displaymath}}
\newcommand{\bi}{\begin{itemize}}
\newcommand{\ei}{\end{itemize}}
\newcommand{\bc}{\begin{center}}
\newcommand{\ec}{\end{center}}
\newcommand{\bfl}{\begin{flushleft}}
\newcommand{\efl}{\end{flushleft}}
\newcommand{\bfr}{\begin{flushright}}
\newcommand{\efr}{\end{flushright}}
\newcommand{\f}{\frac}

\def\bk{{\bf k}} \def\bq{{\bf q}}
  
\def\ra{\rightarrow}
\def\6{\partial} \def\a{\alpha} \def\b{\beta}
 \def\d{\delta} \def\ve{\varepsilon} 
\def\z{\zeta}  \def\th{\theta}
  \def\l{\lambda}
   
  \def\t{\tau}

\def\o{\omega} \def\G{\Gamma} 
  
  \def\O{\Omega}

\def\={\!\!\!&=&\!\!\!}
\def\+{\!\!\!&&\!\!\!+~}
\def\-{\!\!\!&&\!\!\!-~}

\begin{document}

\title{Fluctuation contribution to the specific heat in non-Fermi models 
for superconductivity}
\author{I. Tifrea, I. Grosu and M. Crisan}
\address{Department of Theoretical Physics, University of Cluj, 3400 Cluj, Romania}
\maketitle

\begin{abstract}
We investigate the fluctuation contribution to the specific heat of a two-dimensional 
superconductor with a non-Fermi normal state described by a Anderson Green's function 
$G(\bk,i\o)=\o_c^{-\a}/(i\o-\ve_\bk)^{1-\a}$. The specific heat corrections contain a 
term proportional to $(T^{2\a}-T_c^{2\a})^{-1}$ and another logarithmic one. We 
defined a coherence length as function of the non-Fermi parameter $\a$, which showed that 
a crossover study between BCS and Bose-Einstein condensation is possible by varying $\a$ 
in an interval $0\div \a_{cr}$. By comparing our theoretical results with the experimental 
data for HTSC materials, we reobtained the value for $\a$, corresponding to such systems, 
of the order $0.3\div 0.45$. We also reobtained the critical temperature for such a 
superconductor using the Thouless criterion.
\end{abstract}

\section*{Introduction}
After the discovery of high critical temperature superconductors (HTSC) a great number 
of experimental 
works showed that the usual Fermi liquid description is unappropriated for these materials. As 
a consequence there are several phenomenological models \cite{1,2,3} proposed in order to 
explain the high critical temperature $T_c$ and the nonmetallic behavior of such materials. An 
important role in the mechanism of superconductivity is related to the fluctuation effects which 
are very important in the critical region. One of the parameter that characterize the normal 
state properties of a HTSC is the specific heat. Thompson and Kresin \cite{4}, showed that in 
the absence of a magnetic field ${\bf B}=0$, in the critical region ($T\ra T_c, T>T_c$) the 
corrections in the specific heat obtained due to the fluctuation effects are proportional to 
$(T-T_c)^{-1/2}$. These corrections correspond to the classical region of fluctuation \cite{5,6}. 
Recently Grosu \cite{7} reconsidered the problem for a two-dimensional superconductor with 
a logarithmic density of states and he showed that the corrections due to the fluctuation effects 
in the specific heat consist in two terms, one proportional to $(T-T_c)^{-1}$ and another 
logarithmic one. 

In a previous paper \cite{8} we derived the form for the specific heat jump in a non-Fermi 
superconductor described by the Anderson model. In our calculations we considered that the 
normal state Green's function has a similar form as the one from 1D interacting fermion systems 
\cite{9}. A similar choice for the Green's function $G(\bk,i\o)=\o_c^{-\a}/(i\o-\ve_\bk)^{1-\a}$, 
where $\o_c$ is a frequency cutoff introduced to makes the Green's function dimensionally correct 
and $\a$ is a parameter related to the anomalous Fermi surface, was made by Chakravarty and 
Anderson \cite{10} in order to study the interlayer mechanism of the cuprate-oxides.
The model has been developed by Sudbo \cite{10s} who showed that the parameter
$\a$ has to satisfy $0<\a<0.5$.

The aim of this work is to calculate the specific heat corrections of a two-dimensional non-Fermi 
system using the Anderson model. In Section II we derive the particle-particle polarization for 
the model and we reobtain the critical temperature $T_c$. We introduce using the fluctuation 
propagator a Ginzburg-Landau coherence length extrapolated to zero temperature, which in our 
calculations is function of the parameter $\a$. In Section III we evaluate the correction terms 
in the specific heat due to the fluctuation effects. In Section IV we analyze our results.

\section*{Particle-particle polarization}

In order to evaluate the corrections in the specific heat inside the critical region above $T_c$ 
first we have to calculate the Green's function connected with the Cooper pairs. It was showed by 
Schmid\cite{11} that this Green's function is given by

\be
D(\bq,\o_n)=\left[\f{1}{V}+\Pi(\bq,\o)\right]^{-1}
\label{e1}
\ee
where $V$ is the interaction strength and $\Pi(\bq,\o_n)$ is the usual particle-particle 
polarization given by

\be
\Pi(\bq,\o_m)=T\sum_n\int\f{d^2k}{(2\pi)^2}G(\bk,\o_n)G(\bq-\bk,\o_m-\o_n)
\label{e2}
\ee
with

\be
G(\bk,\o_n)=\f{\o_c^{-\a}}{(i\o_n-\ve_\bk)^{1-\a}}
\label{e3}
\ee
The behavior of the Green function given by Eq. (\ref{e3}) for $\o\ra 0$ has been analyzed
in Ref. \onlinecite{10y} and the anomalous $\o$-dependence of it in the interval
$-\o_c, \o_c$ gives a softer singularity than $1/\o$ specific for the Green function of the
Fermi liquid.

We remark that in the usual Fermi liquid theory ($\a=0$), for a constant density of states, in a 
three dimensional system we reobtained the standard result (See Ref.\onlinecite{11})

\be
D(\bq,\o_n)=\f{1}{N_0[\tau+a|\o_n|+\xi^2q^2]}
\label{e4}
\ee
where $N_0$ is the density of states at the Fermi level, $\tau=(T-T_c)/T_c$, $a=\pi/8T_c$ and 
$\xi$ is the Ginzburg-Landau coherence length extrapolated to the zero temperature.

In order to calculate the particle-particle polarization we will rewrite Eq. (\ref{e2}) in 
the following form

\be
\Pi(\bq,\o_m)=\int\f{d^2k}{(2\pi)^2} S(\bk, \bq, \o_m)
\label{e5}
\ee
where

\be
S(\bk, \bq, \o_m)= 
T\sum_n\f{\o_c^{-2\a}}{(i\o_n-\ve_\bk)^{1-\a}(i\o_m-i\o_n-\ve_{\bq-\bk})^{1-\a}}
\label{e6}
\ee
The evaluation of the sum over the Matsubara frequencies in Eq. (\ref{e6}) is made using a 
contour integral which involves the Fermi-Dirac distribution function, with the specification 
that in this integral we have to deal with two branch cuts related to the complex function 
involved in the summation. Another observation which has to be made is the fact that a correct 
evaluation of this sum is possible only on the assumption that we work at low frequencies $\o_m$ 
and small wavevectors $\bq$. In this way we get

\bea
S(\bk,\bq,\o_m)&=&\f{\o_c^{-2\a}\sin{[\pi (1-\a)]}}{\pi}\nonumber\\
&\times&\left\{-(2\ve_\bk)^{2\a-1}B(1-2\a,\a)\right.\nonumber\\
&+&\f{2\G(\a)}{\sqrt{\pi}}\sum_{m=0}^\infty (-1)^m
\f{(2\ve_\bk)^{\a-1/2}}{[\b(m+1)]^{\a-1/2}}K_{\a-1/2}[\ve_\bk\b(m+1)]\nonumber\\
&+&(\a-1)(v_Fq\cos{\th}-\o_m)\left[-(2\ve_\bk)^{2\a-2}B(2-2\a,\a-1)\right.\nonumber\\
&+&\left.\f{\G(\a-1)}{\sqrt{\pi}}
\sum_{m=0}^\infty (-1)^m\f{(2\ve_\bk)^{\a-1/2}}{[\b(m+1)]^{\a-3/2}}
K_{\a-3/2}[\ve_\bk\b(m+1)]\right]\nonumber\\
&+&\f{(\a-1)^2(v_Fq\cos{\th})^2}{2}\left[-(2\ve_\bk)^{2\a-3}B(3-2\a,a-2)\right.\nonumber\\
&+&\f{\G(\a-2)}{\sqrt{\pi}}\sum_{m=0}^\infty (-1)^m\f{(2\ve_\bk)^{\a-1/2}}{[\b(m+1)]^{\a-5/2}}
K_{\a-3/2}[\ve_\bk\b(m+1)]\nonumber\\
&+&\left.\left.\f{2\G(\a-1)}{\sqrt{\pi}}\sum_{m=0}^\infty (-1)^m\f{(2\ve_\bk)^{\a-3/2}}{[\b(m+1)]^{\a-3/2}}
K_{\a-3/2}[\ve_\bk\b(m+1)]\right]\right\}
\label{e7}
\eea
where for the validity of this formula $\a$ should be between 0 and 0.5, $B(x,y)$ is the beta 
function, $\G(x)$ is the Euler's gamma function and $K_\nu(z)$ is the Bessel function of 
imaginary argument.

Using Eqs. (\ref{e5}) and (\ref{e7}) is very simple to evaluate the 
particle-particle polarization

\bea
\Pi(\bq, \o)&=&\f{2N_0\sin{[\pi(1-\a)]}}{\pi}\nonumber\\
&\times&\left\{-\f{2^{2\a-1}B(1-2\a,\a)}{2\a}\left(\f{\o_D}{\o_c}\right)^{2\a}+
\G^2(\a)\f{1-2^{1-2\a}}{2^{1-2\a}}\z(2\a)\left(\f{T}{\o_c}\right)^{2\a}\right.\nonumber\\
&+&\f{i\o(1-\a)}{\o_c}\left[-\f{2^{2\a-2}B(2-2\a,\a-1)}{2\a-1}
\left(\f{\o_D}{\o_c}\right)^{2\a-1}\right.\nonumber\\
&+&\left.\f{\G(\a-1)\G(\a-1/2)}{\sqrt{\pi}}\f{1-2^{2-2\a}}{2^{2-2\a}}\z(2\a-1)
\left(\f{T}{\o_c}\right)^{2\a-1}\right]\nonumber\\
&+&\f{(v_Fq)^2(1-\a)^2}{4\o_c^2}\left[-\f{2^{2\a-3}B(3-2\a,\a-2)}{2\a-2}
\left(\f{\o_D}{\o_c}\right)^{2\a-2}\right.\nonumber\\
&+&\left.\left.\left(\f{2\G(\a-2)\G(\a-1/2)}{\sqrt{\pi}}+\G^2(\a-1)\right)\f{1-2^{3-2\a}}{2^{3-2\a}}
\z(2-2\a)\left(\f{T}{\o_c}\right)^{2\a-2}\right]\right\}
\label{e8}
\eea
where $\z(x)$ is the Riemann's zeta function.

In the case of zero frequency and zero wavevector $\Pi(\bq=0, i\o=0)$ we can obtain the critical 
temperature for normal state-superconductivity phase transition using the Thouless criterion

\be
1+V Re\Pi(\bq=0, i\o=0)=0
\label{e9}
\ee
It is easy to see that we reobtained the previous result from Ref. \onlinecite{12,13}

\be
T_c^{2\a}=\f{1}{C(\a)}\left[D(\a)\o_D^{2\a}-\f{\o_c^{2\a}}{A(\a)\l}\right]
\label{e10}
\ee
where $\l=N_0V$, $C(\a)=\G^2(\a)(1-2^{1-2\a})\z(2\a)$, $A(\a)=2^{2\a}\sin{[\pi(1-\a)]}/\pi$ and 
$D(\a)=B(1-2\a,\a)/2\a$.

Using Eqs. (\ref{e8}) and (\ref{e10}) the fluctuation propagator can be written in the following 
form

\bea
D^{-1}(\bq, \o)&\cong& N_0A(\a)C(\a)\left[\left(\f{T}{\o_c}\right)^{2\a}-
\left(\f{T_c}{\o_c}\right)^{2\a}\right]\nonumber\\
&+&N_0A(\a)\left[\f{i\o}{T}M(\a)(1-\a)+\f{(v_Fq)^2}{4T^2}N(\a)(1-\a)^2\left(\f{T_c}{\o_c}
\right)^2\right]
\label{e11}
\eea
where

\bea
M(\a)&=&-\f{B(2-2\a,\a-1)}{2(2\a-1)}\left(\f{\o_D}{\o_c}\right)^{2\a-1}\f{T_c}{\o_c}\nonumber\\
&+&\f{\G(\a-1)\G(\a-1/2)}{2\sqrt{\pi}}(1-2^{2-2\a})\z(2\a-1)\left(\f{T_c}{\o_c}\right)^{2\a}
\label{e12}
\eea

\bea
N(\a)&=&-\f{B(3-2\a,\a-2)}{4(2\a-2)}\left(\f{\o_D}{\o_c}\right)^{2\a-2}\nonumber\\
&+&\left(\f{2\G(\a-2)\G(\a-1/2)}{\sqrt{\pi}}+\G^2(\a-1)\right)\f{1-2^{3-2\a}}{4}\z(2-2\a)
\left(\f{T_c}{\o_c}\right)^{2\a-2}
\label{e13}
\eea
which can be written also in the following form

\be
D(\bq, \o)=\f{1}{N_0\left[b\t+ia\o+\xi_0^2(\a)q^2\right]}
\label{e14}
\ee
with $a=(1-\a)M(\a)A(\a)/T$, $b\cong A(\a)C(\a)(T_c/\o_c)^2$, $\t=(T/T_c)^{2\a}-1$ and

\be
\xi_0^2(\a)\cong \f{v_F^2(1-\a)^2}{4\o_c^2}N(\a)A(\a)
\label{e15}
\ee

\begin{figure}
\centering
\includegraphics[clip,width=0.8\textwidth]{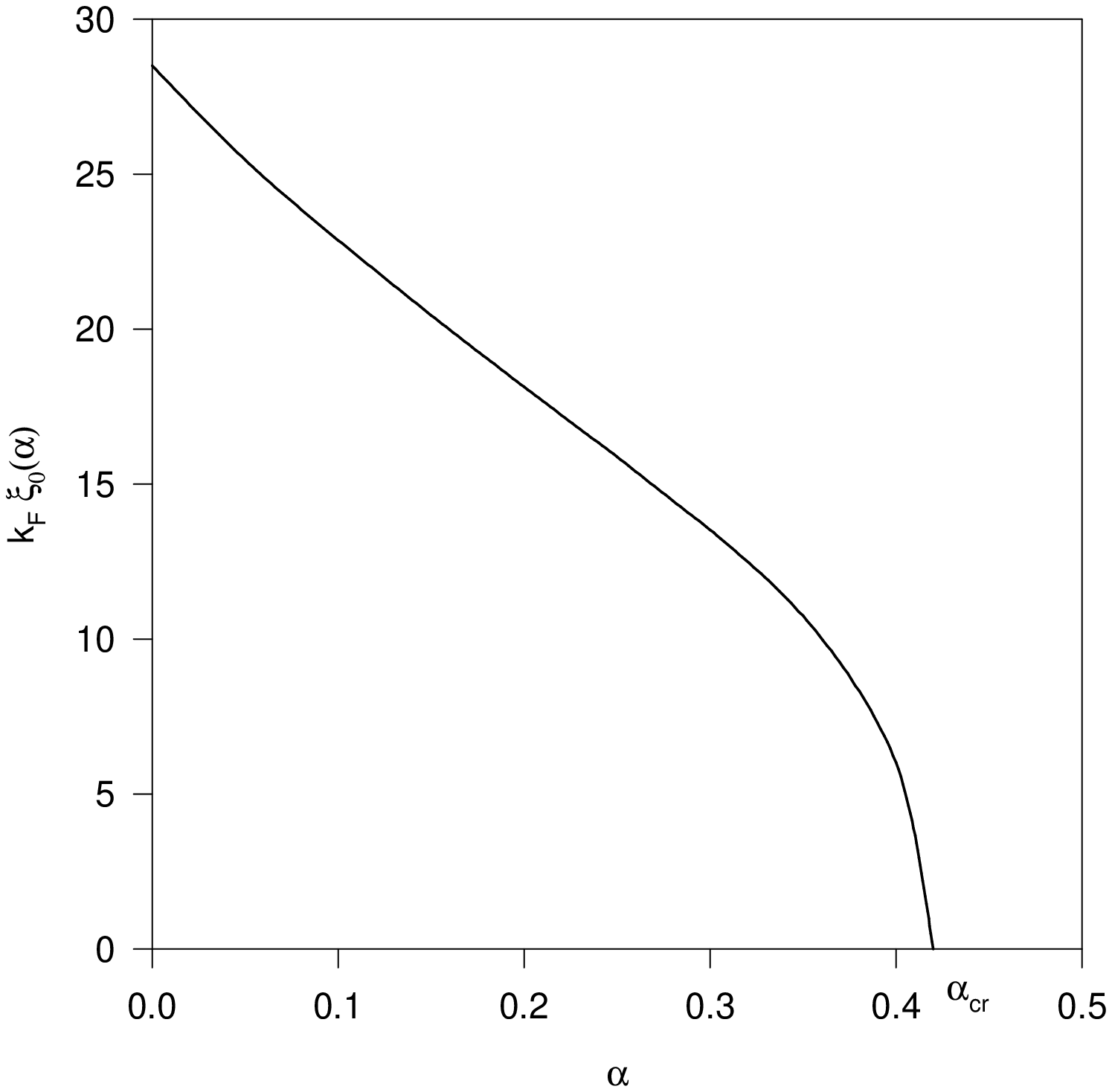}
\caption{The parameter $k_F\xi_0(\a)$ as a function of the non-Fermi 
parameter $\a$ ($\o_D/\o_c=0.2$, $E_c/\o_c=30$, $T_c=89 K$).}
\label{fig1}
\end{figure}

If we compare Eqs. (\ref{e14}) and (\ref{e4}) we can define a 
Ginzburg-Landau coherence length extrapolated to zero temperature, which now is a 
function of the non-Fermi parameter $\a$. As it is known an important parameter related to the 
crossover theory between a BCS like formalism and a Bose-Einstein condensation 
formalism\cite{14} is $k_F\xi_0$. In the BCS limit the value of such a parameter is about 
10$^3$$\div$10$^4$ and in the Bose-Enstein condensation limit is zero. Experimental data showed 
that in the case of HTSC materials this parameter is between this two limits, for example in 
the case of YBa$_2$Cu$_3$O$_{7-\d}$ is about 5$\div$10. In Fig. (\ref{fig1}) we plot the variation 
of the parameter $k_F\xi_0$ (with the critical temperature corresponding to the optimal doping of 
YBa$_2$Cu$_3$O$_{7-\d}$) as a function of the non-Fermi parameter $\a$. The limit $\a\ra 0$ 
correspond to the overdoped limit where the usual Fermi liquid theory still works for the 
normal state properties. The other limit $\a\ra \a_{cr}$ correspond to the underdoped limit where 
the usual Fermi liquid theory breaks down. In this limit as we can see from Fig. (\ref{fig1}) 
the coherence length is going to zero, which shows that here a Bose-Einstein condensation 
of strong coupling pairs ($\xi\ra 0$) is more close to the real picture of the system. The 
intermediate limit ($\a\ra 10$), which correspond to the optimal doping of 
YBa$_2$Cu$_3$O$_{7-\d}$, shows that the non-Fermi parameter $\a$ should be around 0.30$\div$0.45, 
result which is in agreement with our previous calculation of the specific heat jump at the 
critical temperature\cite{8}.


\section*{Fluctuation Contribution to the Specific Heat}

The specific heat corrections due to the fluctuations around the critical temperature, will be 
calculated by having an additional term in the thermodynamic potential. Inside the critical region 
only the zero frequency term is important 

\be
\O_{fl}=T\sum_\bq\ln{\left[D^{-1}(\bq,i\o=0)\right]}
\label{e16}
\ee
Using the well known thermodynamic relations $S=-\6\O_{fl}/\6 T$ and $C=T\6 S/\6 T$ the specific 
heat correction close to $T_c$ is 

\be
C_{fl}\cong -T_c\f{\6^2\O_{fl}}{\6\t^2}\left(\f{\6\t}{\6 T}\right)^2-T_c
\f{\6\O_{fl}}{\6\t}\f{\6^2\t}{\6 T^2}
\label{e17}
\ee
From Eqs. (\ref{e14}) and (\ref{e17}) we get

\be
C_{fl}\cong \f{b\a^2}{\pi\xi_0^2(\a)}\f{T_c^{2\a}}{T^{2\a}-T_c^{2\a}}+
\f{b\a(1-2\a)}{2\pi\xi_0^2(\a)}
\ln{\left(\f{2mE_F\xi_0^2(\a)}{b}\f{T_c^{2\a}}{T^{2\a}-T_c^{2\a}}\right)}
\label{e18}
\ee
The result contained in Eq. (\ref{e18}) is valid only in the critical region when we consider 
the classical fluctuations (region where the random phase approximation is still valid). Close 
to the transition point the interaction between electrons and Cooper pairs are very important, 
so we expect that in some way the electronic Green's function is affected. As a consequence there 
is a contribution $\d\Pi$ to the particle-particle polarization $\Pi$ in Eq. (\ref{e2}), which 
can be written as

\be
\d\Pi=2T^2\sum_n\int\f{d^2p}{(2\pi)^2}G^2(\bk,\o_n)G^2(-\bk,-\o_n)\int\f{d^2q}{(2\pi)^2}
D(\bq,i\o=0)
\label{e19}
\ee
If we perform the integrals over $\bk$ and $\bq$, and if we sum over the Matsubara frequencies 
$\o_n$ we get

\be
\d\Pi\cong\f{B(1/2,2(1-\a)-1/2)}{4\pi^3}\f{\o_c^{-4\a}}{(\pi T_c)^{1-4\a}}
\f{2^{3-4\a}-1}{2^{3-4\a}}\z(3-4\a)\ln{\f{2mE_F\xi_0^2(\a)}{b\t}}
\label{e20}
\ee
With this correction the particle-particle polarization becomes $\Pi_t=\Pi+\d\Pi$ and the 
fluctuation propagator can be written as

\be
D_t^{-1}=D^{-1}+\d\Pi
\label{e21}
\ee
Using this definition very close to the transition temperature the significant contribution 
due to the fluctuations will be 

\be
\O_t\cong g\ln{\left|c\ln{\f{2mE_F\xi_0^2(\a)}{b\t}}\right|}
\label{e22}
\ee
where $g$ is a constant which contains the critical temperature $T_c$ and a cutoff wavevector 
of order of $q_F$ (the Fermi wavevector) and $c$ is the constant in front of the logarithmic 
function in Eq. (\ref{e20}). With this thermodynamic potential the fluctuation contribution to 
the specific heat close to the critical temperature becomes

\be
C_t\cong \f{4\a^2 g}{T_c}\left[\t^2\ln{\f{2mE_F\xi_0^2(\a)}{b\t}}\right]^{-1}+
\f{2\a(2\a-1)g}{T_c}\left[\t\ln{\f{2mE_F\xi_0^2(\a)}{b\t}}\right]^{-1}
\label{e23}
\ee

As we can see from Eqs. (\ref{e23}) and (\ref{e18}) the effect of the interaction between 
electrons and fluctuations should be very important, at least close enough to $T_c$. The 
only question is if the formula from Eq. (\ref{e23}) is valid in the framework of the performed 
approximations. There are several observations 
which make questionable this formula. The most important one is the fact that in Eq. (\ref{e19}) 
we used instead of a renormalized fluctuation propagator the bare one, which gives us a 
different behavior as a function of the wavevector $\bq$ for $\d\Pi$. In fact Eq. (\ref{e19}) 
should be a self consistent equation for $\d\Pi$, which in our work is solved in the zero order.

\section*{Discussion}
We evaluated the contribution of the fluctuation contributions in the specific heat of a 
non-Fermi system described by the Anderson model\cite{1}. We derived the form of the 
particle-particle polarization, and we showed that the critical temperature can be reobtained 
using the Thouless criterion. Using the fluctuation propagator $D(\bq,i\o)$ we introduced a 
coherence length $\xi_0(\a)$, which inn our calculations is function of the non-Fermi parameter 
$\a$. By varying $\a$ between zero and $\a_{cr}$ we showed in the case of YBa$_2$Cu$_3$O$_{7-\d}$ 
that the system can be described using the BCS theory or by the Bose-Einstein condensation 
of preformed pairs. In the case of optimal doping we reobtained for $\a$ the same values as in 
our previous paper\cite{8} $\a=0.3\div 0.45$. As a consequence we consider that the numerical 
analysis in Ref.\onlinecite{12,13} should be made on a scale $\a=0.3\div 0.45$, not on the 
scale very close to the value $\a=0$. 

In the third section of our paper we calculated the fluctuation contribution in the specific 
heat close to the critical temperature $T_c$, related to the classical fluctuation region, where 
the random phase approximation still remains valid. The result obtained by considering the 
fluctuation contributions give two different terms, the significant one being proportional to 
$(T^{2\a}-T_c^{2\a})^{-1}$. A similar result can be obtained if we consider the critical field 
$H_{c2}(\a,T)$ obtained by Muthukumar et al.\cite{12} using the Ginzburg-Landau theory. Close to 
the critical temperature $T_c$, where interaction between electrons and Cooper pairs becomes 
very important we calculated the behavior of the specific heat by introducing a correction in the 
particle-particle polarization. The corrections obtained due to this additional term is very 
important, the divergence in Eq. (\ref{e23}) being more powerfull than the one in Eq. 
(\ref{e18}). All these results are valid in the approximation of the "box corrections" 
for the fluctuations propagator.

Finally we mention that the superconducting state developed in
Ref. \onlinecite{10y} using this non-Fermi model presents some
differences to the results from
Ref. \onlinecite{10s} because of the phase factor in the Green function.
However, using the model proposed in Ref. \onlinecite{10s}
in the limit $\a\ra 0$ it was showed \cite{13} that the critical temperature is
identical to the BCS expression.

\end{document}